\documentclass[preprint, amsfonts]{revtex4}
\usepackage{amsfonts}
\usepackage{amsmath}
\usepackage{bm}
\usepackage{graphicx}
\usepackage{epsf}

\def\prd{ Phys. Rev. D }
\def\pra{ Phys. Rev. A }
\def\prl{ Phys. Rev. Lett. }
\def\pr{ Phys. Rev. }
\def\pla{ Phys. Lett. A }

\begin{document}
\title{Three-region vacuum nonlocality}

\author{ Jonathan Silman }
\email{silmanjo@post.tau.ac.il}
\author{ Benni Reznik }
\email{reznik@post.tau.ac.il}
\affiliation{ School of Physics and Astronomy, Raymond and Beverly Sackler
Faculty of Exact Sciences,
Tel-Aviv University, Tel Aviv 69978, Israel. }
\date{January 6, 2005}
\begin{abstract}
\bigskip

The correlations between three arbitrarily far-apart regions of the
vacuum state of the free Klein-Gordon field are investigated by means
of its finite duration coupling to three localized detectors. It is
shown that these correlations cannot be reproduced in terms of a hybrid
local-nonlocal hidden-variable model, i.e., the correlations between
three arbitrarily separated regions of the vacuum are fully nonlocal.
\end{abstract}
\maketitle
In a recent paper \cite{Reznik}, the nature of the correlations between
two arbitrarily far-apart regions of the ground state of the free
Klein-Gordon field was investigated by means of its finite duration
coupling to a pair of localized detectors. It was shown that a local
hidden-variable (HV) model cannot account for these correlations \cite{Reznik, Werner},
and that as a function of the separation between the regions, $L$,
and the duration of the coupling, $T$, the entanglement decreases
at a slower rate than $e^{-\left(L/cT\right)^{3}}$. It is, therefore,
natural to ask whether the vacuum admits other types of these kind
of correlations, i.e. \emph{true} multi-region entanglement \cite{Acin},
and \emph{full} nonlocality \cite{Svetlichny}. In this paper we will
answer both questions affirmatively for the case of three arbitrarily
far-apart regions of the vacuum. We will follow the method developed
in previous papers \cite{Reznik,Reznik2}.

Our method consists of the finite duration coupling of the field we
wish to investigate to any number of localized nonentangled detectors,
such that all the detectors remain causally disconnected from one
another throughout the interaction. Once the interaction is over,
we trace over the field degrees of freedom to obtain the detectors'
reduced density matrix. The crux of the method lies in the fact that
since the detectors are initially nonentangled, any nonlocal correlations
exhibited by the detectors' final reduced density matrix must have
their origin in the vacuum. This enables us to apply recently developed
tools from the field of quantum information theory to study the structure
of the vacuum.

Before we begin, let us first give the definitions of true multi-partite
entanglement and full nonlocality. A multi-partite mixed state is
said to be truly multi-partite entangled iff it cannot be expressed
as a convex sum of decomposable terms. In the tri-partite case this
just means that the state cannot be written as a convex sum of terms
of the form $\rho_{i}\otimes\rho_{jk}$, where the subscripts denote
any of the three subsystems. Examples of truly tri-partite entangled
states are the GHZ \cite{GHZ} and W \cite{W} states. Analogously,
we can also distinguish between several types of nonlocality. A multi-partite state is fully nonlocal
 if there does not exist a decomposable hidden-variable dependent probability  function that can account for
the results of any von Neumann measurement. In the tri-partite case of a system composed of three parts $A$, $B$, and $C$, this means that
$\wp_{ABC}\left( a, \, b, \, c \mid \lambda \right)\neq \wp_{A}\left( a \mid \lambda\right)\wp_{BC}\left( b, \, c \mid \lambda \right)+
\wp_{C}\left( c \mid \lambda\right)\wp_{AB}\left( a, \, b \mid \lambda \right)+
\wp_{B}\left( b \mid \lambda\right)\wp_{CA}\left( c, \, a \mid \lambda \right)$
. Here $\lambda$ is the hidden-variable, and
$\wp_{ABC}\left( a, \, b, \, c \mid \lambda \right)$ is the probability for $\hat{a}=a$, $\hat{b}=b$, and $\hat{c}=c$.
 Otherwise, the state may admit a hybrid local-nonlocal hidden-variable description \cite{Footnote0}.
 Svetlichny derived a Bell-like inequality to distinguish
between these two cases \cite{Svetlichny}.\\

Let us consider the ground state of a free Klein-Gordon field and
three nonentangled point-like two-level detectors \cite{Footnote}.
The interaction Hamiltonian of the field and the detectors is given
by
\begin{eqnarray}
H_{I}\left(t\right)&=&H_{I}^{A}\left(t\right)+H_{I}^{B}\left(t\right)+H_{I}^{C}\left(t\right)\\
&=&\sum_{i=A,\, B,\, C}\int_{-T/2}^{t}dt'\epsilon_{i}\left(t'\right)\left(e^{i\Omega_{i}t}\sigma_{i}^{+}+e^{-i\Omega_{i}t}\sigma_{i}^{-}\right)
\phi\left(\vec{x}_{i},\, t'\right)\nonumber,
\end{eqnarray}
where $\phi\left(\vec{x},\, t\right)$ is a free Klein-Gordon field
in three spatial dimensions, the $\sigma_{i}^{\pm}$ are the detectors'
{}``ladder'' operators, and the $\Omega_{i}$ denote the energy
gap between detector energy levels. $T$ is the duration of the interaction,
while the window-functions, $\epsilon_{i}\left(t\right)$, govern
its strength. We set $cT<<L_{ij}$, with $L_{ij}\equiv\left|\vec{x}_{i}-\vec{x}_{j}\right|$
and the $\vec{x}_{i}$ being the locations of the detectors. This
ensures that the detectors remain causally disconnected throughout
the interaction. The evolution operator therefore factors to a product.
 In the Dirac interaction representation, employing
 ``natural'' units ($\hbar=c=1$), $U=\prod_{i=A,\, B,\, C}\hat{T}e^{-i\int dtH_{I}^{i}\left(t\right)}$,
with $\hat{T}$ denoting time ordering. Expanding to the square order
in the $\epsilon_{i}\left(t\right)$, once the interaction is over the final state of the
system is given by\begin{eqnarray}
U\left|0\right\rangle \left|\downarrow\downarrow\downarrow\right\rangle  & \simeq & \left|0\right\rangle \left|\downarrow\downarrow\downarrow\right\rangle -i\Phi_{A}^{+}\left|0\right\rangle \left|\uparrow\downarrow\downarrow\right\rangle -i\Phi_{B}^{+}\left|0\right\rangle \left|\downarrow\uparrow\downarrow\right\rangle -i\Phi_{C}^{+}\left|0\right\rangle \left|\downarrow\downarrow\uparrow\right\rangle \nonumber \\
 &  & -\Phi_{A}^{+}\Phi_{B}^{+}\left|0\right\rangle \left|\uparrow\uparrow\downarrow\right\rangle -\Phi_{B}^{+}\Phi_{C}^{+}\left|0\right\rangle \left|\downarrow\uparrow\uparrow\right\rangle -\Phi_{C}^{+}\Phi_{A}^{+}\left|0\right\rangle \left|\uparrow\downarrow\uparrow\right\rangle \\
 &  & -\sum_{i=A,\, B,\, C}\Theta_{i}\left|0\right\rangle \left|\downarrow\downarrow\downarrow\right\rangle +i\Phi_{A}^{+}\Phi_{B}^{+}\Phi_{C}^{+}\left|0\right\rangle \left|\uparrow\uparrow\uparrow\right\rangle +O\left(\epsilon^{3}\right),\nonumber\end{eqnarray}
where $\Phi_{i}^{\pm}\equiv\int_{\scriptscriptstyle -T/2}^{\scriptscriptstyle T/2}dt\epsilon_{i}\left(t\right)e^{\pm i\Omega_{i}t}\phi\left(\vec{x}_{i},\, t\right)$,
and $\Theta_{i}\equiv\frac{1}{2}\hat{T}\left[\int_{\scriptscriptstyle -T/2}^{\scriptscriptstyle T/2}dtH_{I}^{i}\left(t\right)\int_{\scriptscriptstyle -T/2}^{\scriptscriptstyle T/2}dt'H_{I}^{i}\left(t'\right)\right]$.
(Actually the last term in the expansion is of cubic order, because
unlike the other cubic terms, it cannot simply be discarded at this
stage.) When working in the computational basis, \{$\downarrow\downarrow\downarrow,\, \downarrow\downarrow\uparrow,\,\downarrow\uparrow\downarrow,\,\downarrow\uparrow\uparrow,\,\uparrow\downarrow\downarrow,\,
\uparrow\downarrow\uparrow,\,\uparrow\uparrow\downarrow,\,\uparrow\uparrow\uparrow$\},
the detectors' nonnormalized reduced density matrix is given by 
\begin{widetext}\begin{equation}
{\footnotesize
\left(\begin{array}{cccccccc}
1-C & 0 & 0 & -d_{BC}^{++} & 0 & -d_{CA}^{++} & -d_{AB}^{++} & 0\\
0 & d_{CC}^{-+} & d_{BC}^{-+} & 0 & d_{CA}^{-+} & 0 & 0 & d_{ABCC}^{---+}\\
0 & d_{BC}^{-+} & d_{BB}^{-+} & 0 & d_{AB}^{-+} & 0 & 0 & d_{ABCB}^{---+}\\
-d_{BC}^{++} & 0 & 0 & d_{BCBC}^{--++} & 0 & d_{CABC}^{--++} & \eta^{2}d_{ABBC}^{--++} & 0\\
0 & d_{CA}^{-+} & d_{AB}^{-+} & 0 & d_{AA}^{-+} & 0 & 0 & d_{ABCA}^{---+}\\
-d_{CA}^{++} & 0 & 0 & d_{CABC}^{--++} & 0 & d_{CACA}^{--++} & d_{ABCA}^{--++} & 0\\
-d_{AB}^{++} & 0 & 0 & d_{ABBC}^{--++} & 0 & d_{ABCA}^{--++} & d_{ABAB}^{--++} & 0\\
0 & d_{ABCC}^{---+} & d_{ABCB}^{---+} & 0 & d_{ABCA}^{---+} & 0 & 0 & d_{ABCABC}^{---+++}\end{array}\right)+O\left(\epsilon^{4}\right)}.
\end{equation}\end{widetext}
\vspace{1.5em}
Here we have employed the notation $d_{i\cdots n}^{\alpha\cdots\zeta}\equiv\left\langle 0\right|\Phi_{i}^{\alpha}\cdots\Phi_{n}^{\zeta}\left|0\right\rangle $,
and $C\equiv\sum_{i}\left\langle \downarrow\downarrow\downarrow\right|\left\langle 0\right|\Theta_{i}\left|0\right\rangle \left|\downarrow\downarrow\downarrow\right\rangle $,
with $i,\,\ldots,\, n=A,\, B,\, C$ and $\alpha,\,\ldots,\,\zeta=\pm$.
For simplicity, we have chosen temporally symmetric window-functions.
Hence the amplitudes are all real. $d_{ii}^{-+}$ is the amplitude
for a single photon emission by detector $i$, while $d_{i,\, j\neq i}^{++}$
is the amplitude for a single virtual photon exchange between detectors
$i$ and $j$. The physical interpretation of the rest of the amplitudes
should thus be clear.

To prove that a multi-partite mixed state does not admit a hybrid
local-nonlocal HV description, it is enough to show that it can be
distilled to a fully nonlocal pure state \cite{Popescu,Gisin}.
 Having each of the detectors
pass through a filter, which attenuates its {}``spin-down'' component
by a factor of $\eta$, the detectors' nonnormalized reduced density
matrix is in the computational basis given by 
\begin{widetext}\begin{equation}{\footnotesize
\left(\begin{array}{cccccccc}
\eta^{6} & 0 & 0 & -\eta^{4}d_{BC}^{++} & 0 & -\eta^{4}d_{CA}^{++} & -\eta^{4}d_{AB}^{++} & 0\\
0 & \eta^{4}d_{CC}^{-+} & \eta^{4}d_{BC}^{-+} & 0 & \eta^{4}d_{CA}^{-+} & 0 & 0 & \eta^{4}d_{ABCC}^{---+}\\
0 & \eta^{4}d_{BC}^{-+} & \eta^{4}d_{BB}^{-+} & 0 & \eta^{4}d_{AB}^{-+} & 0 & 0 & \eta^{4}d_{ABCB}^{---+}\\
-\eta^{4}d_{BC}^{++} & 0 & 0 & \eta^{2}d_{BCBC}^{--++} & 0 & \eta^{2}d_{CABC}^{--++} & \eta^{2}d_{ABBC}^{--++} & 0\\
0 & \eta^{4}d_{CA}^{-+} & \eta^{4}d_{AB}^{-+} & 0 & \eta^{4}d_{AA}^{-+} & 0 & 0 & \eta^{4}d_{ABCA}^{---+}\\
-\eta^{4}d_{CA}^{++} & 0 & 0 & \eta^{2}d_{CABC}^{--++} & 0 & \eta^{2}d_{CACA}^{--++} & \eta^{2}d_{ABCA}^{--++} & 0\\
-\eta^{4}d_{AB}^{++} & 0 & 0 & \eta^{2}d_{ABBC}^{--++} & 0 & \eta^{2}d_{ABCA}^{--++} & \eta^{2}d_{ABAB}^{--++} & 0\\
0 & \eta^{4}d_{ABCC}^{---+} & \eta^{4}d_{ABCB}^{---+} & 0 & \eta^{4}d_{ABCA}^{---+} & 0 & 0 & d_{ABCABC}^{---+++}\end{array}\right)}.
\end{equation}
\end{widetext}
\vspace{1.5em}
Note that each of the components is written to its lowest nonvanishing
order. The reason for this will shortly become apparent.
For $L_{ij}>>T$, the overlap amplitudes, $d_{i,\, j\neq i}^{-+}$,
are negligible as compared to the emission, $d_{ii}^{-+}$, and exchange
amplitudes, $d_{i,\, j\neq i}^{++}$.  If we now
take the window-function of detector $C$, and only detector $C$,
to be superoscillatory \cite{Aharonov,Berry}, of a form as in \cite{Reznik3,Reznik}, then by a suitable
choice of the remaining two window-functions and the $L_{ij}$ the exchange amplitudes involving
the superoscillatory window-function can be made arbitrarily larger
than the rest \cite{Reznik, Footnote2}, i.e.,  $d_{BC}^{++}=d_{CA}^{++}\gg all\, other\, amplitudes$.
 In this limit, if we set $\eta^{2}=d_{BC}^{++}=d_{CA}^{++}$,
the detectors' reduced density matrix is just \begin{widetext}\begin{equation}
{\footnotesize \frac{1}{3} \left(\begin{array}{cccccccc}
1 & 0 & 0 & -1 & 0 & -1 & 0 & 0\\
0 & 0 & 0 & 0 & 0 & 0 & 0 & 0\\
0 & 0 & 0 & 0 & 0 & 0 & 0 & 0\\
-1 & 0 & 0 & 1 & 0 & 1 & 0 & 0\\
0 & 0 & 0 & 0 & 0 & 0 & 0 & 0\\
-1 & 0 & 0 & 1 & 0 & 1 & 0 & 0\\
0 & 0 & 0 & 0 & 0 & 0 & 0 & 0\\
0 & 0 & 0 & 0 & 0 & 0 & 0 & 0\end{array}\right)}.\end{equation}\end{widetext}
\vspace{1.5em}
This density matrix is pure and corresponds to the state $\frac{1}{\sqrt{3}}\left(\left|\downarrow\downarrow\downarrow\right\rangle -\left|\downarrow\uparrow\uparrow\right\rangle -\left|\uparrow\downarrow\uparrow\right\rangle \right)$,
which, by means of local operations on each of the detectors,
can be transformed into a W state, $\frac{1}{\sqrt{3}}\left(\left|\uparrow\downarrow\downarrow\right\rangle +\left|\downarrow\uparrow\downarrow\right\rangle \right.$
$\left.+\left|\downarrow\downarrow\uparrow\right\rangle \right)$.
The W state violates the Svetlichny inequality \cite{Cereceda}, and
is therefore fully nonlocal. Since the detectors are initially nonentangled
and remain causally disconnected throughout the interaction, we conclude
that these correlations must have their origin in corresponding vacuum
correlations, i.e., the correlations between three arbitrarily separated
regions of the vacuum are fully nonlocal. And since true multi-partite
entanglement is a necessary condition for full nonlocality, it immediately
follows that the vacuum is truly tri-partite entangled as well.
\begin{acknowledgments}
We acknowledge support from the ISF (Grant No. 62/01-1).
\end{acknowledgments}


\begin{thebibliography}{10}
\bibitem{Footnote0}The fact that a decomposable hidden-variable dependent probability function accounting for all von Neumann measurements can be found, does not
exclude the presence of ``hidden'' full nonlocality. See \cite{Popescu, Gisin}.
\bibitem{Footnote}In \cite{Reznik}, the correlations between two non-overlapping regions
of the vacuum were studied using spatially {}``smeared'' relativistic
detectors. Note that identical results to those obtained there, can
be obtained, following exactly the same protocol, using nonrelativistic
point-like detectors. The same is true here.
\end{thebibliography}
\end{document}